\documentclass[12pt]{article}
\usepackage{geometry}             
\usepackage{graphicx}
\usepackage{amssymb}
\usepackage{amsmath}
\usepackage{epstopdf}
\usepackage{comment}

\usepackage[hyperindex=true,
          pdfstartview=FitH,
          bookmarksnumbered=true,
          bookmarksopen=true,
          citecolor=blue,
          linkcolor=blue,
          colorlinks=true,
          pdfborder=001,
          unicode]{hyperref}

\allowdisplaybreaks

\begin{document}

\title{Near horizon symmetry and entropy of black holes in
the presence of a conformally coupled scalar
}
\author{Kun Meng${}^{1}$, Zhan-Ning Hu${}^{1}$ and Liu Zhao${}^{2}$\\
${}^{1}$ School of Science, Tianjin Polytechnic University, 
Tianjin 300387, China\\
${}^{2}$ School of Physics, Nankai University, Tianjin 300071, China\\
{\em email}: \href{mailto:mengkun@tjpu.edu.cn}{mengkun@tjpu.edu.cn}, 
\href{mailto:zhanninghu@ya​hoo.com.cn}{zhanninghu@ya​hoo.com.cn} and \\
\href{mailto:lzhao@nankai.edu.cn}{lzhao@nankai.edu.cn}}
\date{}                      
\maketitle

\begin{abstract}
We analyze the near horizon conformal symmetry for black hole solutions in  
gravity with a conformally coupled scalar field using the method proposed by 
Majhi and Padmanabhan recently. It is shown that the entropy of the black holes 
of the form $\mathrm{d}s^2 = - f(r)\mathrm{d}t^2 + \mathrm{d}r^2/f(r)+...$ 
agrees with Wald entropy.  This result is 
different from previous result obtained by M. Natsuume, T. Okamura and M. 
Sato using the canonical Hamiltonian formalism, which claims a discrepancy from 
Wald entropy.
\end{abstract}

\section{Introduction}
Holographic duality is one of the most interesting subject in high energy 
physics since 't Hooft and Susskind\cite{tHooft, Susskind} proposed the so-called 
holographic principle. The most well-known realization of holographic duality is
through AdS/CFT. Though the idea of AdS/CFT becomes popular only after 
Maldacena's celebrated work \cite{Maldacena}, the first known example of 
CFT dual of AdS gravity was actually constructed before Maldacena's work. 
Early in 1986, Brown and Henneaux found the conformal symmetry on the 
asymptotic boundary of AdS spacetime using the canonical Hamiltonian formalism 
\cite{Brown}. Using this result, Strominger gave an explanation of the 
microscopic origin of $(2+1)$-dimensional black hole entropy \cite{Strominger} 
with the aid of Cardy's formula \cite{Cardy}, and the result agrees with  
Bekenstein-Hawking entropy. Instead of working at the asymptotic infinity, 
Carlip extended such studies to the near horizon region of black hole solutions 
\cite{Carlip1,Carlip2}, and also obtained Bekenstein-Hawking entropy through 
Cardy's formula. Subsequent works are too numerous to be listed here, though it 
is worth mentioning the recent development in the studies of Kerr/CFT 
correspondence, which generalizes the result to axisymmetric spacetimes 
\cite{StromingerKerrCFT,Lu,Carlip3}. Most of the works mentioned above take the 
view point that the near horizon conformal symmetry is the consequence of 
the asymptotic symmetry of the total action of the system, i.e. bulk plus 
boundary contribution.

Recently, Majhi and Padmanabhan \cite{Padmanabhan} proposed a new method 
(MP approach for short) for constructing the near horizon conformal symmetry. 
The Noether current they constructed is related to 
diffeomorphism invariance of the {\em boundary action} alone, as apposed to the  
whole theory. This different construction is reasonable, because it is natural
to expect that the entropy of the black hole is originated only from the 
degrees of freedom around or on the relevant null surface. The Killing vectors 
that leave the boundary action invariance under diffeomorphism is chosen 
to keep the form of the metric near the horizon in order to ensure that
the horizon is not destroyed by the diffeomorphism.  A direct calculation shows 
that the algebra of the conserved 
charges is Virasoro algebra, whose central charge can be inserted into 
Cardy‘s formula to get the black hole entropy. For pure gravity without matter 
source, the resulting entropy matches the Bekenstein-Hawking entropy.

A natural question one wants to ask is if the quantum nature of a black hole is 
also captured by a dual conformal field theory when matter source come in, the 
simplest example being the black hole solutions in gravitational theories 
coupled with a scalar field. For minimally coupled scalar,
existing studies show that the central charge is the same as the pure gravity 
\cite{0201170}. As to the conformally coupled case, through the canonical 
Hamiltonian formalism proposed by Brown and Henneaux, it was shown 
\cite{Natsuume} that the {\em central charge} thus obtained is the same as in 
the pure gravity case, consequently the entropy 
{\em disagrees} with the Wald entropy \cite{Wald}. 
This is a quite weird situation, because it is generally believed that the Wald 
entropy is universal for gravitational theories with higher curvature and/or 
matter source contribution. In this paper, we are aimed to reconsider the
near horizon symmetry and entropy of black holes in the presence of a 
conformally coupled scalar field using the MP approach. It turns out that the
near horizon symmetry still closes a Virasoro algebra, however, the central 
charge receives a contribution from the scalar field and is different from that
one would expect in pure gravitational theories or in the case with a minimally 
coupled scalar field. However, the entropy obtained using Cardy's formula does 
agree with the Wald entropy.

This paper is organized as follows: In section 2, we briefly review the MP 
approach for constructing the Noether current arising from invariance of the
boundary action, taking pure Einstein gravity as an example. 
In section 3, we study the conformal symmetry in 
the near horizon region of a special class of black hole solution of the form
(\ref{staticbh1}) in the presence of a conformally coupled scalar field and 
calculate the black hole entropy using Cardy's formula. It turns out that the 
Virasoro central charge is different from the minimally coupled case. However, 
the entropy matches the Wald entropy. The last section of the paper is devoted 
to the summarization of the results.

\section{Boundary action and MP approach}

The MP approach relies purely on the {\em boundary action} of the gravitational 
theory under consideration. As is well known, the variational process of
a gravitational theory requires that there is a boundary counter term which 
cancels out the total divergence term arising from the variation of the bulk
Lagrangian. This boundary counter term is what is referred to as the boundary 
action in this paper. 

According to Stokes theorem, any boundary action can be rewritten as a bulk 
integral. Omitting the integration symbol, we can identify the integrand as
\begin{align}
\sqrt{g}L=\sqrt{g}\nabla_a A^a,\label{Lag}
\end{align}
where $A^a$ is a vector field defined on the bulk spacetime. Under a 
diffeomorphism $x^a\rightarrow x^a+\xi^a(x)$, the variation of the left hand 
side of (\ref{Lag}) is given by
\begin{equation}
\delta_{\xi}\left(\sqrt{g}L\right)
\equiv\mathcal{L}_{\xi}\left(\sqrt{g}L\right)
=\sqrt{g}\nabla_a\left(L\xi^a\right),\label{varileft}
\end{equation}
where $\mathcal{L}_{\xi}$ denotes the Lie derivative along the vector field $\xi^a(x)$ and we have $\mathcal{L}_{\xi}\left(\sqrt{g}\right)=
\sqrt{g}\nabla_a\xi^a$ and $\mathcal{L}_{\xi}(L)=\xi^a\nabla_aL$.
Similarly, the right hand side changes by
\begin{align}
&\delta_{\xi}\left(\sqrt{g}\nabla_aA^a\right)
=\mathcal{L}_{\xi}\left[\partial_a
\left(\sqrt{g}A^a\right)\right]\nonumber\\
&\quad=\partial_a\left[A^a\mathcal{L}_{\xi}
\sqrt{g}+\sqrt{g}\mathcal{L}_{\xi}A^a\right]\nonumber\\
&\quad=\sqrt{g}\nabla_a\left[\nabla_b\left(A^a\xi^b\right)
-A^b\nabla_b\xi^a\right], \label{varright}
\end{align}
where $\nabla_aA^a=\frac{1}{\sqrt{g}}\partial_a
\left(\sqrt{g}A^a\right)$ is used.

Equating eq.(\ref{varileft}) and eq.(\ref{varright}), we get the conservation of 
the following Noether current,
\begin{align*}
  J^a\left[\xi\right]=L\xi^a-\nabla_b
  \left(A^a\xi^b\right)+A^b\nabla_b\xi^a.
\end{align*}
Inserting (\ref{Lag}) into this formula, we get
\begin{align}
  J^a\left[\xi\right]=\nabla_b J^{ab}[\xi],
\end{align}
where 
\begin{align}
J^{ab}[\xi]=\nabla_b\left[\xi^aA^b-\xi^bA^a\right]
\end{align}
is known as the {\em Noether potential}. As we have seen, the Noether current
$J^a[\xi]$ arises purely from variations of the the boundary action rather than 
from the bulk one.

For pure Einstein gravity without matter source, the boundary action is given by 
the well-known York-Gibbons-Hawking term
\begin{align}
I_B&=\frac{1}{8\pi G}\int_{\partial\mathcal{M}}
  \mathrm{d}^{d}x\sqrt{\sigma}K\nonumber\\
  &=\frac{1}{8\pi G}\int_\mathcal{M}\mathrm{d}^{d+1}x\sqrt{g}
  \nabla_a\left(KN^a\right),
\end{align}
where $N^a$ is the unit vector field normal to the boundary hypersurface 
$\partial\mathcal{M}$ and $K=-\nabla_a N^a$ is the trace of the extrinsic 
curvature. The Noether current is now given by
\begin{align}
  J^a\left[\xi\right]=\nabla_aJ^{ab}=
  \frac{1}{8\pi G}\nabla_b\left(K\xi^a N^b-K\xi^b N^a\right).\label{current}
\end{align}
The corresponding Noether charge is defined as
\begin{align}
  Q\left[\xi\right]=\frac{1}{2}
  \int_{\partial\Sigma}\sqrt{h}\,\mathrm{d}
  \Sigma_{ab}\,J^{ab},\label{charge}
\end{align}
where $\mathrm{d}\Sigma_{ab}=-\mathrm{d}^{(d-1)}x
\left(N_aM_b-N_bM_a\right)$ is the area element on the $(d-1)$-dimensional 
hypersurface $\partial\Sigma$. $N^a$ and $M^a$
are respectively spacelike and timelike unit normal vectors.

Finally, the algebra of the conserved charges is defined as:
\begin{align}
  &\left[Q_1,Q_2\right]\equiv\frac{1}{2}
  \left(\delta_{\xi_1}Q[\xi_a]-\delta_{\xi_a}
  Q[\xi_1]\right)\nonumber\\
  &=\frac{1}{2}\int_{\partial\Sigma}\sqrt{h}
  \mathrm{d}\Sigma_{ab}\left[\xi_2^aJ^b[\xi_1]
  -\xi_1^aJ^b[\xi_2]\right],\label{commutator}
\end{align}
where $\delta_{\xi_1}Q[\xi_2]=\int_{\Sigma}
\mathrm{d}\Sigma_a\,\mathcal{L}_{\xi_1}\left(
\sqrt{g}J^a[\xi_2]\right)$. As was shown in \cite{Padmanabhan}, this algebra 
leads to the Virasoro algebra in the near horizon limit. In the next section, we 
shall show that similar constructions also works for a special class
of black hole solutions in gravitational theories with a non-minimally coupled 
scalar field.

\section{Near horizon symmetry in the presence of a 
conformally coupled scalar field}

In this section, we consider the near horizon symmetries of a static black hole
with a conformally coupled scalar hair. The desired Virasoro algebra can be 
derived using the MP approach, from which the entropy is given with
Cardy's formula. The result agrees with Wald entropy as will be seen.

The bulk action of gravity with a conformally coupled scalar field reads
\begin{align}
I=\int_\mathcal{M}\mathrm{d}^{d+1}x\sqrt{-g}\left[\frac{1}{16\pi G}
\left({R-\frac{d(d-1)}{\ell^2}}\right)
-\frac{1}{2}g^{\mu\nu}\partial_\mu\phi\partial_\nu\phi
-\frac{d-1}{8d}R\phi^2-U(\phi)\right],\label{bulkaction}
\end{align}
where $\ell$ is the AdS radius and $U(\phi)$ is the scalar potential. 
The scalar field $\phi$ couples to gravity not only through the kinetic term but 
also through the $R\phi^2$ term. With the special value $\frac{d-1}{8d}$ of the
curvature coupling constant, the kinetic term for the scalar field and the
curvature coupling term put together are invariant under conformal rescaling
of the metric field, that is why this special value of curvature coupling is 
referred to as conformal coupling. The boundary 
term associated with the above action is \cite{0703152}
\begin{align}
  I_B=\frac{1}{8\pi G}\int_{\partial\mathcal{M}}\mathrm{d}^dx\sqrt{h}
  \left(1-\frac{d-1}{d}2\pi G\phi^2\right)K.\label{ourboundary}
\end{align}
In deriving the above boundary term, we take the usual
boundary conditions that keep the metric and scalar field fixed but not their 
normal derivatives, i.e. $\delta g_{\mu\nu}=\delta\phi=0$ on 
$\partial\mathcal{M}$, while $n^\rho\partial_\rho\delta g_{\mu\nu}$ and 
$n^\rho\partial_\rho\delta \phi$ do not vanish. From the bulk action 
(\ref{bulkaction}), we see that, only the kinetic term contributes to
the boundary term after taking variation, which is proportional to 
$\delta\phi n^\mu\nabla_\mu\phi$, thus we do not need a boundary term to cancel 
it. The variation with respect to the scalar field of the boundary term 
(\ref{ourboundary}) also vanishes. As to the variation of the metric, everything 
goes like the case of pure gravity with only
an additional factor $\left(1-\frac{d-1}{4d}8\pi G\phi^2\right)$. As a 
confirmation, one can take a conformal transformation that transforms the 
non-minimally coupled case to the minimal case, then the boundary term becomes 
the standard York-Gibbons-Hawking term, as 
can be seen in \cite{0703152,0809.4033,0107077}.

Now we consider a static black hole metric of the form
\begin{align}
 \mathrm{d}s^2=-f(r)\mathrm{d}t^2+\frac{\mathrm{d}r^2}{f(r)}
 +r^2\mathrm{d}\Omega^2,  \label{staticbh1}
\end{align}
where $\mathrm{d}\Omega^2$ is the line element on a unit $(d-1)$-sphere. 
Though not being the most general form of static, spherically symmetric metrics, 
the above form of metric ansatz is very frequently adopted in
pure gravity (e.g. in Schwarzschild metric), gravity with minimally coupled 
matter sources (see, e.g. \cite{referee1}) as well as in gravity with 
nonminimally coupled 
sources \cite{Martinez:1996p2505,Nadalini:2007p2561,XZ}. Note that the black hole 
metric given in \cite{Martinez:1996p2505} is just the one used in 
\cite{Natsuume}, which led to an entropy which is different from Wald entropy 
using the canonical Hamiltonian formalism.

Suppose that there is an event horizon at $r=r_h$, i.e. $f(r_h)=0$.  In order to 
consider the near-horizon region, it is convenient for us
to introduce a new coordinate $\rho$ via $r= \rho+r_h$, in terms of which the 
metric becomes
\begin{align}
 \mathrm{d}s^2=-f(\rho+r_h)\mathrm{d}t^2+
 \frac{\mathrm{d}\rho^2}{f(\rho+r_h)}+(\rho+r_h)^2\mathrm{d}
 \Omega^2\label{staticbh}
\end{align}
In the near horizon approximation, i.e. $\rho\rightarrow0$, $f(\rho+r_h)$ can be 
expanded as $f(\rho+r_h)=2\kappa \rho+\frac{1}{2}f^{\prime\prime}
(r_h)\rho^2+\cdots$, with $\kappa=\frac{f^{\prime}(r_h)}{2}$ being the surface 
gravity. In the leading order approximation, the well known  Rindler  metric can 
be achieved $\mathrm{d}s^2=-2\kappa\rho\mathrm{d}t^2+\frac{1}{2\kappa\rho}
\mathrm{d}\rho^2+\cdots$.

In order to obtain a Virasoro algebra, the vector field $\xi^\mu$ should be 
taken properly. It is convenient for us to transform to the Bondi-like metric
to solve the Killing equations and then transform back. Using the coordinate 
transformation
\begin{align}
 \mathrm{d}u=\mathrm{d}t-
 \frac{\mathrm{d}\rho}{f(\rho+r_h)},
\end{align}
the metric (\ref{staticbh}) can be rewritten as
\begin{align}
 \mathrm{d}s^2=-f(\rho+r_h)\mathrm{d}u^2
 -2\mathrm{d}u\mathrm{d}\rho+(\rho+r_h)^2\mathrm{d}
 \Omega^2.
\end{align}

Now, we impose the horizon-structure-keeping conditions
\begin{align}
&\mathcal{L}_\xi g_{\rho\rho}=-2\partial_\rho\xi^u=0, \nonumber\\
&\mathcal{L}_\xi g_{u\rho}=-f(\rho+r_h)\partial_\rho\xi^u-
\partial_\rho\xi^\rho -\partial_u\xi^u=0.
\end{align}
Solving the above Killing equations we obtain
\begin{align}
\xi^u=F(u,x),\ \ \ \xi^\rho=-\rho\partial_uF(u,x),
\end{align}
where $x$ denotes the coordinates on the unit $(n-1)$-sphere. All other 
components of $\xi^\mu$ vanish. The condition $\mathcal{L}_\xi g_{uu}= 0$ is 
satisfied in the near-horizon limit after a direct calculation. Now, we return 
to the original coordinates ($t,\rho$), the vector fields take the form
\begin{align}
  \xi^t=T-\frac{\rho}{f(\rho+r_h)}\partial_tT,
  \ \ \xi^\rho=-\rho\partial_t T,\label{outxi}
\end{align}
where $T(t,\rho, x)\equiv F(u,x)$.

For the metric (\ref{staticbh1}), the normal vectors can be chosen as
\begin{align}
  N^\mu=\left(0, \sqrt{f(r)},0,0,\cdots\right),\ \ M^\mu=\left(\frac{1}
  {\sqrt{f(r)}},0,0,0,\cdots\right).\label{normalvector}
\end{align}
Substituting eqs.(\ref{outxi}) and (\ref{normalvector}) into eq.(\ref{charge}),
and inserting the boundary action (\ref{ourboundary}), the Noether charge can 
be given as
\begin{align}
Q[\xi]=\frac{1}{8\pi G}\left(1-\frac{d-1}{d}2\pi G\phi^2(r_h)\right)
  \int_{\partial\Sigma}\mathrm{d}^{d-1}x\sqrt{h}
  \left(\kappa T-\frac{1}{2}\partial_t T\right).\label{charge2}
\end{align}

Note that once $T$ is determined, the vector field and Noether charge are 
determined, and they are both linear in $T$. We can expand $T$ in terms of a set 
of basis functions
\begin{align}
  T=\sum_mA_mT_m,
\end{align}
where $A_m^{*}=A_{-m}$ in order  that $T$ is real.
Since to every $T_m$ there is a corresponding $\xi^\mu_m$, the basis functions 
$T_m$ should be chosen properly in order that the Diff $S^1$ algebra is 
satisfied:
\begin{align}
i\left\{\xi_m,\xi_n\right\}^\mu=(m-n)\xi^\mu_{m+n},
\end{align}
where $\{,\}$ is the Lie bracket. The correct form of $T_m$ is \cite{0204179}
\begin{align}
  T_m=\frac{1}{\alpha}\exp\left[im\left(\alpha
  t+g(\rho)+p\cdot x\right)\right],\label{basis}
\end{align}
where $\alpha$ is a constant, $p$ is an integer , $g(\rho)$ is a regular 
function on the horizon. The expanded modes of $Q[\xi]$ corresponding to 
$T_m$ will henceforth be denoted as $Q_m$. 
The commutators between these modes can be calculated straightforwardly 
by substituting eqs. (\ref{current}), (\ref{outxi}), (\ref{normalvector}) into  
(\ref{commutator}),
\begin{align}
\left[Q_m,Q_n\right]
&=\frac{1}{8\pi G}\left(1-\frac{d-1}{d}2\pi G\phi^2(r_h)\right)
  \int_{\partial\Sigma}\mathrm{d}^{d-1}x\sqrt{h}\bigg[
  \kappa(T_m\partial_tT_n-T_n\partial_tT_m)\nonumber\\
  &-\frac{1}{2}(T_m\partial^2_tT_n-T_n\partial^2_tT_m)
  +\frac{1}{4\kappa}(\partial_tT_m\partial^2_tT_n-\partial_tT_n\partial^2_tT_m)
  \bigg].
\end{align}
Performing the integration, the final form of the modes $Q_m$ and the 
commutators are obtained explicitly,
\begin{align}
Q_m&=\frac{1}{8\pi G}\left({1-\frac{d-1}{d}2\pi G\phi^2(r_h)}\right)
\frac{\kappa A}{\alpha}\delta_{m,0},\nonumber\\
\left[Q_m,Q_n\right]
&=\frac{1}{8\pi G}\left({1-\frac{d-1}{d}2\pi G\phi^2(r_h)}\right)
\left[-\frac{i\kappa A}{\alpha}(m-n)\delta_{m+n,0}-im^3\frac{\alpha A}{2\kappa}
\delta_{m+n,0}\right],
\end{align}
where $A$ is the area of the horizon of the black hole.

Then the central term of the Virasoro algebra is
\begin{align}
K\left[\xi_m, \xi_n\right]&=\left[Q_m, Q_n\right]+i(m-n)Q_{m+n}\nonumber\\
&=-im^3\frac{\alpha A}{2\kappa}\frac{1-\frac{d-1}{4d}8\pi G\phi^2(r_h)}{8\pi G}
\delta_{m+n,0}
\end{align}
The central charge and zero mode energy $Q_0$ can be read off easily
\begin{align}
 \frac{C}{12}&=\frac{A}{16\pi G}\frac{\alpha}{\kappa}\left(1-\frac{d-1}{d}2\pi G
 \phi^2(r_h)\right),\\
 Q_0&=\frac{A}{8\pi G}\frac{\kappa}{\alpha}\left(1-\frac{d-1}{d}2\pi G
 \phi^2(r_h)\right).
\end{align}
Using Cardy's formula, we obtain the entropy of the black hole
\begin{align}
  S=2\pi\sqrt{\frac{CQ_0}{6}}=\frac{A}{4G}\left(1-\frac{d-1}{d}2\pi G\phi^2(r_h)
  \right).\label{entropy}
\end{align}
which is exactly the Wald entropy. This result is different from the previous 
work \cite{Natsuume}, which adopted the Brown-Henneaux canonical formalism. In 
\cite{Natsuume}, under the black hole no hair assumption, Natsuume, Okamura and 
Sato obtained the same central charge as in pure gravity theories, thus the 
black hole entropy obtained from the central charge does not agree with the Wald 
entropy. However, using the MP approach, we showed that the scalar field does 
contributes to the central charge of the near horizon Virasoro algebra and 
the Wald entropy is recovered.

\section{Conclusion}

In this paper, we extended the MP approach for constructing near horizon 
conformal symmetries for pure Einstein gravity theory to the case of gravity 
with a conformally coupled scalar field.
The work presented in this paper confirms that the entropy of black holes in the 
presence of conformally coupled scalar field is identical to Wald entropy as 
apposed to previous claim for a discrepancy. Also, our construction 
does not depend on the concrete choice of the black hole metric (i.e. the 
function $f(r)$ remain unspecified), therefore the MP approach can be applied to 
a broader class of black holes at once, as apposed to the traditional 
Hamiltonian approach, which depends sensitively on the concrete choice of the 
black hole solution as well as the near-horizon behavior of each metric 
elements. We believe that our work is helpful in further extending the 
application of the MP approach as well as in further understanding the 
universality of Wald entropy.

\providecommand{\href}[2]{#2}\begingroup
\footnotesize\itemsep=0pt
\providecommand{\eprint}[2][]{\href{http://arxiv.org/abs/#2}{arXiv:#2}}

\end{document}